\documentstyle[preprint,floats,aps,epsf,epsfig,prb]{revtex}
\tightenlines
\newlength{\bxwidth}\bxwidth=2.5 truein
\newcommand\om { \omega_n}

\newlength{\fight}
\newcommand\ltdash{\raise-1.8pt\hbox{$\scriptscriptstyle |$}}
\newcommand \beq  {\begin{equation}}
\newcommand \eeq  {\end{equation}}
\newcommand \bea {\begin{eqnarray} }
\newcommand \eea {\end{eqnarray}}
\newcommand \up{\uparrow}
\newcommand \dw{\downarrow}

\newcommand \ra { \rangle}
\newcommand\la {\langle}

\newcommand\al{\alpha}
\newcommand\be{\beta}
\newcommand \bk { {\bf k}}
\newcommand \ek { \varepsilon_{\bk}}
\newcommand \dk {\Delta_{\bk}}
\newcommand \bR { {\bf R}}
\newcommand \bq { {\bf q}}

\begin{document}
\draft
%***********    This is for two columns *******************************
%\twocolumn[\hsize\textwidth\columnwidth\hsize\csname @twocolumnfalse\endcsname
%********************************

\title{
Density of States of a d-wave Superconductor in the Presence of Strong
Impurity Scatterers: a Non Perturbative Result
}
\author{ Catherine P\'epin $^{1,2}$ and Patrick A. Lee
$^{2}$}

\address{$1$ Department of Physics, 
Oxford University, 
1 Keble Road, 
Oxford OX1 3NP,UK }

\address{$2$ Department of Physics, Massachussetts Institute of
Technology, Cambridge, MA 02139, USA }

\maketitle
\date{\today}
\maketitle
%\widetext
\begin{abstract}
We present a method to compute the density of states induced by a
finite density of non
magnetic impurities in a d-wave superconductor, in the unitary limit
of very strong scattering centers. For frequencies very small as
compared to the superconducting gap ($\omega \ll
\Delta_0$) the additional density of states has the leading divergence
${\displaystyle \delta \rho (\omega) \simeq  n_i \; / \left ( | 2 \omega | \ln^2
|\omega/\Delta_0|  \right ) }$. This result is non perturbative. 
\end{abstract}
\vskip 0.2 truein
\newpage
% ***********    This is for two columns *******************************
%\vskip2pc]
%********************************
As a consequence of the breakdown of Anderson's
theorem~\cite{anderson}, when impurity scattering violates the
symmetry of the superconducting state, the superconducting energy gap
is depleted and  impurities act as strong pair breakers. This is the
case in s-wave superconductors with magnetic
impurities~\cite{abrikosov,kadanoff} for which there is creation of
bound states in the gap. Since they are not in the same symmetry
representation, non magnetic impurities act as pair breakers in
unconventional superconductors with higher orbital momentum such as
d-wave superconductors. As a result, the measured low temperature
properties of YBaCu$_2$0$_7$~\cite{hirsh1} displays a remarkable
sensitivity to the presence of non magnetic impurities: the critical
temperature, for example, is suppressed even for very small density
of impurities~\cite{alloul}.
Furthermore the d-wave superconductor is special, due to the presence of gap nodes which prevents the complete freezing of scattering processes at low energy.

\par
The standard method to treat these problems of disorder~\cite{joynt7,tsv2,rama,joynt} combines the T-matrix approximation with standard impurity averaging techniques. For three dimensional systems such as polar superconductors or heavy Fermion superconductors~\cite{joynt6} the standard perturbative approach is reliable. In the limit of low impurity concentration $n_i$, a perturbative expansion in $n_i$ leads to a finite density of states at the chemical potential~\cite{tsv2,joynt12}.

\par
For two dimensional systems (to which it is believed the high-T$_C$
cuprates belong) the standard procedure of averaging over impurities
may be complicated by the appearance of logarithmic singularities in
the perturbative expansion of the single electron
self-energy~\cite{alexei}. Such a situation appears in a variety of
two dimensional systems characterized by a Dirac-like canonical
spectrum. This remark has cast some doubt upon the validity of
perturbative expansions in $n_i$ because at each loop-level
logarithmic divergences prevent the series to converge. Nevertheless
self-consistent versions~\cite{hatsug,dhlee,hettler} of the standard averaging
technique have been performed showing a finite density of states at
zero energy. On the other hand, some non perturbative methods have
been used to treat a weak disorder
potential~\cite{alexei,mudry,rephettler,senthil}. The solution then
depends on the symmetry of the pure system. In the special case of the
d-wave superconductor the density of states still vanishes at low
energy in the presence of disorder. Recently Ref.~\onlinecite{senthil}
concluded that the density of states is finite above a very low energy
scale (essentially the level spacing of a localization volume), below
which a pseudo-gap appears.
Early numerical simulations~\cite{xiang} seem to confirm finite
density of states at zero  energy.
\par
The issue of finite density of states is crucial for the conduction
properties in the disordered compound. Indeed, if there exist states
in the gap, possible anomalous overlaps between well separated
impurities can induce a new conduction mechanism entirely through
impurity wave functions in a so-called `impurity
band'~\cite{balatsky,patrick} and may lead to delocalization.
On the other hand, the results of Ref.~\onlinecite{senthil} 
concluded that the states are localized. 

\par
 Here, we reexamine the issue of the density of states
in a dirty d-wave superconductor. We consider the limiting case of
a dilute concentration $n_i$ of identical impurities in the unitary limit. 
In other words, each impurity is a strong 
s-wave scatterer which is represented by an infinitely strong
point-like repulsive potential (infinite scattering potential $V_0$)
whose position is random. Using standard perturbative techniques,
previous studies concluded that the density of states was finite
at the Fermi energy \cite{joynt7,rama,joynt,hatsug,dhlee}.
In contrast, using non-perturbative techniques, we find that the density of
states is singular at the Fermi energy,
$ 
\rho (\omega) = 
n_i / \left (2| \omega| \ln^2 \left | \omega / \Delta_0 \right | \right )
$.
The origin of this singular density of states is the existence
of impurity bound states at the Fermi energy ($E_F=0$). It has been shown~\cite{autres1,khaliullin} that one single impurity in the unitary limit creates a bound state at $E=0$, with a spatial envelope that decays as $1/ (R \ln R)$. For many impurities these states overlap, but our result indicates that a singular density of states
remains at zero energy of the form  $ \rho (\omega) = n_i / \left ( 2
 | \omega| \ln^2 \left | \omega / \Delta_0 \right | \right )$.

In this paper, we introduce a new method to calculate the leading
divergence in the density of states induced by $N$ non magnetic impurities in a system of two dimensional Dirac fermions in the unitary limit. 
We will outline the general features of the proof and present a different derivation than in our previous work~\cite{cath}.
The question of averaging will be discussed in details and we show how the result is exact provide the average over the different configurations of impurities is performed at the end of the calculation.
Our result can be applied to several systems of Dirac fermions in two dimensions. We explicitly treat the d-wave superconductor.

The paper is organized as follows. In section I we introduce the model of BCS d-wave superconductor, diagonalize it and define notations. In section II we establish the T-matrix equation starting from the equations of motion. We also introduce the matrix ${\hat M}$ as the inverse of the T-matrix. In section III we prove a sum-rule useful to calculate the density of states. Section IV is dedicated to a detailed study of the structure and matrix elements of ${\hat M}$. Section V contains the heart of the proof. We first discuss the effect of taking the unitary limit. Then we indicate which are the quantities where the divergence appears, leading to the singular density of state that we found. In the conclusion we discuss how our result relates to the different theories of disordered superconductors.

\section{The model}

The generic Hamiltonian for a d-wave superconductor can be written 
\beq
\label{eq115}
H_0= \sum_{ {\bf k } } \phi_{ {\bf k } }^\dagger 
\left[ \ek  \ \sigma_3 + \dk \sigma_1 \right ] \phi_{ {\bf k } }.
\eeq
 It describes BCS quasiparticles with the kinetic energy
 $\ek= W \left( \cos{k_x} + \cos{k_y} \right) -\mu$
 ($\mu$ is the chemical potential)
 in the presence of the spin singlet superconducting order parameter 
 $\dk = \Delta_0 \left( \cos{k_x} -\cos{k_y} \right)$. 
 Distances are measured in units of the
 lattice constant. The $\sigma_i$ are the Pauli matrices in
 the particle-hole space 
\[ \begin{array}{cc}
 \sigma_1 = \left ( \begin{array}{cc} 0 & 1 \\
                       1 & 0 \end{array} \right ) \ , \; \; \; \; \;  & 

  \sigma_3 = \left ( \begin{array}{cc} 1 & 0 \\
                       0 & -1 \end{array} \right ) 
   \end{array} \ . \]
The spinor  
$\phi_{ {\bf k } }^\dagger = \left( c_{\bk,\up}^\dagger, c_{-{ {\bf
                       k}}, \dw} \right)$
creates a particle and a hole with momenta ${\bf k}$ and $-{\bf k}$,
respectively. We shall present our results in terms of a
$d_{x^2-y^2}$ state even though our conclusions apply more generally
to any state where $\dk$ vanishes 
linearly along a direction parallel to the Fermi surface.

\par

Instead of using the Nambu formalism, we work with the diagonalized
version of~(\ref{eq115}) in order to access directly the properties
of quasiparticles. The Bogoliubov transformation that diagonalizes
$H_0$ is given by
\bea
c_{{\bf k} \up} & = & u_{ {\bf k } }
\alpha_{ {\bf k } } - v_{ {\bf k } } \beta_{ {\bf k } } \ ,
\quad
\al_{ {\bf k}}  =  u_{ {\bf k}}  c_{{\bf k} \up} + v_{ {\bf k}}
c^\dagger_{-{ {\bf k}} \dw} \ , 
\nonumber \\
c^\dagger_{-{ {\bf k}} \dw} & = & v_{ {\bf k}} \alpha_{ {\bf k}} +u_{
{\bf k}} \beta_{ {\bf k}} \ ,
\quad
\be_{ {\bf k }} = - v_{ {\bf k}} c_{{\bf k} \up} + u_{ {\bf k}}
c^\dagger_{-{ {\bf k}} \dw},
\eea
where $\alpha_{ {\bf k}}$ and $\beta_{ {\bf k}}$ 
create a particle and a hole with momentum $k$ and
the coefficients $u_{ {\bf k}}$
and $v_{ {\bf k}}$ satisfy 
\beq
\begin{array}{l}
u_{ {\bf k}}^2 = 1/2 \left( 1+ {\displaystyle \frac{\ek}{ \omega_{\bk}} }
\right) \ , \\
v_{ {\bf k}}^2 = 1/2 \left( 1- {\displaystyle \frac{\ek}{\omega_{\bk}} }
\right) \ , \\
u_{ {\bf k}} v_{ {\bf k}}= {\displaystyle  \frac{\dk}{2 \omega_{\bk}}
} \ ,
\end{array} 
\eeq 
with ${\displaystyle \omega_{\bk} = \sqrt{ \ek^2 + \dk^2} }$. 
Given the short-hand notation
\beq
\begin{array}{l}
\psi^\dagger_{ \bk, 0 } \equiv \al^\dagger_\bk \ , \\
\psi^\dagger_{\bk, 1} \equiv \be^\dagger_\bk \ ,
\end{array}
\eeq 
the BCS Hamiltonian can now be rewritten 
\[
H_0 = \sum_{\bk}\sum_{\nu=0,1} \omega_{\bk} (-1)^\nu \psi^\dagger_{\bk,\nu}
\psi_{\bk,\nu}  \ . \]  
The disorder is introduced through $N$ repulsive scalar
potentials $V_0$ located at random positions in the lattice:
\beq
H_I = 
V_0 \sum_{i=1}^N \sum_{\sigma=\uparrow,\downarrow} 
c^\dagger_{i \sigma} c_{i \sigma} \ .
\eeq
The full BCS Hamiltonian $ H= H_0 + H_I$ describes a 
dirty d-wave superconductor.

\section{The T-matrix equation}

\subsection{The Hamiltonian}

With the help of a Fourier transformation to the reciprocal
lattice, the impurity potential becomes
\bea
H_I & =  \frac{V_0}{{\cal V}} \sum\limits_i 
\sum\limits_{\bk, \bk^\prime} e^{i (\bk
  -\bk^\prime) \cdot \bR_i} \left ( c^\dagger_{\bk \up} c_{\bk^\prime
  \up} + c^\dagger_{\bk \dw} c_{\bk^\prime \dw} \right ) \ ,
\eea where ${ \cal V}$ is the volume of the system. 
Rewriting the impurity term in terms of quasiparticles gives
\[
c^\dagger_{\bk \up} c_{\bk^\prime
  \up} = u_\bk u_{\bk^\prime} \al^\dagger_\bk \al_{\bk^\prime} - v_\bk
  u_{\bk^\prime} \be^\dagger_\bk \al_{\bk^\prime} - u_\bk
  v_{\bk^\prime} \al^\dagger_\bk \be_{\bk^\prime} + v_\bk
  v_{\bk^\prime} \be^\dagger_\bk \be_{\bk^\prime} \ .
\]
Thus
\beq
c^\dagger_{\bk \up} c_{\bk^\prime
  \up} = \sum_{\nu,\nu^\prime=0}^1 
  (-1)^\nu (-1)^{\nu^\prime} t_{\bk \nu} t_{\bk^\prime \nu^\prime}
  \psi^\dagger_{\bk \nu} \psi_{\bk^\prime \nu^\prime} \ ,
\eeq 
where we have introduced the short-hand notation
\beq
\begin{array}{cc}
\begin{array}{l}
t_{\bk, 0} \equiv u_\bk \ , \\
t_{\bk, 1} \equiv v_\bk \ .
 \end{array}  \end{array}
\eeq Similarly,
\[
c^\dagger_{-\bk \dw} c_{-\bk^\prime
  \dw} = 
- \left ( v_\bk v_{\bk^\prime} \al^\dagger_\bk \al_{\bk^\prime} + v_\bk
  u_{\bk^\prime} \be^\dagger_\bk \al_{\bk^\prime} + u_\bk
  v_{\bk^\prime} \al^\dagger_\bk \be_{\bk^\prime} + u_\bk
  u_{\bk^\prime} \be^\dagger_\bk \be_{\bk^\prime} \right ) + const.\ ,
\] 
and, neglecting the constant term, we get
\beq
c^\dagger_{-\bk \dw} c_{-\bk^\prime
  \dw} = - \sum_{\nu,\nu^\prime}  t_{\bk \nu +1} t_{\bk^\prime \nu^\prime+1}
  \psi^\dagger_{\bk \nu} \psi_{\bk^\prime \nu^\prime} \ .
\eeq
In summary, the random BCS Hamiltonian can be written
\bea
H & = & \sum_{\bk , \nu} \omega_{\bk} (-1)^\nu \psi^\dagger_{\bk,\nu}
\psi_{\bk , \nu} \nonumber \\
& + & \frac{V_0}{{\cal V}} \sum_i \sum_{\bk, \bk^\prime, \nu, \nu^\prime} e^{i (\bk
  -\bk^\prime) \cdot \bR_i} \left [ (-1)^\nu (-1)^{\nu^\prime} t_{\bk \nu}
  t_{\bk^\prime \nu^\prime} - t_{\bk \nu+1} t_{\bk^\prime \nu^\prime+1}
  \right ] \psi^\dagger_{\bk \nu} \psi_{\bk^\prime \nu^\prime} \ .
\eea

\subsection{Equations of motion}

As the impurities break translation invariance,
the anomalous two-point function
\beq
G_{\bk \bq}^{\nu\nu^\prime} (\tau ) =\left  \la T_\tau \left [
\psi_{\bk \nu} (\tau) \psi^{\dagger}_{\bq \nu^\prime} (0) \right ] 
\right \ra  
\label{eq: anomalous Green function}
\eeq
depends on two momenta.
The equations of motion are
\beq
\label{mo}
- \sum_{\bq, m} {\cal L}_{\bk  \bq}^{\nu m } G_{\bq \bk^\prime}^{m
 \nu^\prime} = \delta_{\bk \bk^\prime} \delta_{\nu 
\nu^\prime} \ ,
\eeq 
where $ {\cal L}_{\bk \bk^\prime}^{\nu \nu^\prime } $ is
\bea
\label{lag}
{\cal L}_{\bk \bk^\prime}^{\nu \nu^\prime } & = &  
[\partial_\tau +(-1)^\nu \omega_{\bk} ]
\delta_{\bk \bk^\prime} \delta_{\nu \nu^\prime}
\nonumber \\
& + & \frac{V_0}{{\cal V}} \sum_i  e^{i (\bk
  -\bk^\prime) \cdot \bR_i}  (-1)^\nu (-1)^{\nu^\prime} t_{\bk \nu}
  t_{\bk^\prime \nu^\prime} \nonumber \\
 & - & \frac{V_0}{{\cal V}} \sum_i  e^{i (\bk
  -\bk^\prime) \cdot \bR_i} t_{\bk \nu+1} t_{\bk^\prime \nu^\prime+1} \ .
\eea 
  We are dealing with a problem of non-interacting particles
  scattered by the static potential generated by $N$ impurities.
  As we shall see, the anomalous Green function in 
  Eq.~(\ref{eq: anomalous Green function}) 
  can be solved by inverting a $2 N \times 2 N$ matrix. 
  To this end, define the one-point functions
\beq
G^0_{\bk \nu}(\tau) \equiv \frac{-1}{\partial_\tau + (-1)^\nu \omega_{\bk} } \
,
\eeq
\beq
\begin{array}{l}
g^1_{\bk \nu}(\bR_i) \equiv (-1)^\nu e^{-i \bk \cdot \bR_i} t_{\bk
\nu} G^0_{\bk 
\nu} \ , \\ g^2_{\bk \nu} (\bR_i) \equiv e^{-i \bk \cdot \bR_i} t_{\bk \nu+1}
G^0_{\bk \nu}\, , 
\end{array}
\eeq
together with
\bea
A_{ij} & =  & \int \frac{d^2 k}{(2 \pi)^2} e^{-i \bk \cdot (\bR_i -\bR_j)}
(t_{\bk \nu} )^2 G^0_{\bk \nu} \ , \\
C_{ij} & = & \int \frac{d^2 k}{(2 \pi)^2} e^{-i \bk \cdot (\bR_i -\bR_j)}
(t_{\bk \nu +1} )^2 G^0_{\bk \nu} \ ,\\
B_{ij} & = & \int \frac{d^2 k}{(2 \pi)^2} e^{-i \bk \cdot (\bR_i -\bR_j)}
(-1)^n t_{\bk \nu} t_{\bk \nu +1}  G^0_{\bk \nu} \ .
\eea 
The integration over $\bk$ is to be performed over the first Brillouin
zone. Equation (\ref{lag}) is inserted into the
equations of motion~(\ref{mo}). We then solve a $2 N \times 2 N$ system of
linear equations (see appendix~\ref{ap1}). This gives
\beq
G_{\bk \bk^\prime}^{\nu \nu^\prime} = G^0_{\bk \bk^\prime} -
\frac{V_0}{{\cal V}} \; \ {\bf N}^T_{- \bk  \nu} \cdot {\hat M}^{-1} \cdot {\bf
N}_{\bk^\prime \nu^\prime} \ ,
\eeq
with ${\bf N}_{\bk \nu}$ a vector made of the $2 N$  components
\beq
\begin{array}{lll}
{\bf N}_{\bk \nu} \equiv \left ( \begin{array}{c} {\bf N}^1_{\bk \nu}
\  \\
{\bf N}^2_{\bk \nu} \end{array} \right ) \ , \; \; \; & {\bf N}^1_{\bk \nu}
\equiv \left ( \begin{array}{c} g^1_{\bk \nu} (\bR_1) \\ \vdots \\
g^1_{\bk \nu}(\bR_N) \end{array} \right ) \ , \; \; \; &  {\bf N}^2_{\bk \nu}
\equiv \left ( \begin{array}{c} g^2_{\bk \nu} (\bR_1) \\ \vdots \\
g^2_{\bk \nu}(\bR_N) \end{array} \right ) \ ,
\end{array}
\eeq 
and ${\hat M}$ is a $2N \times 2 N$ matrix defined by 
\beq
\label{m}
{\hat M} = \left [ \begin{array}{cc}
                  -{\hat I} + V_0 {\hat A} & V_0 {\hat B} \\
                  V_0 {\hat B} & {\hat I} + V_0 {\hat C}
                 \end{array} \right ] \ ,
\eeq with ${\hat A}$, ${\hat B}$, ${\hat C}$ are $N \times N$ matrices whose
matrix elements are respectively $A_{ij}$, $B_{ij}$ and
$C_{ij}$. ${\hat I}$ is the identity matrix.
With these definitions the T-matrix equation can be written
\beq
G_{\bk \bk^\prime}^{\nu \nu^\prime} = G^0_{\bk \bk^\prime} + {\bf N}^T_{-
\bk  \nu} {\hat T} {\bf
N}_{\bk^\prime \nu^\prime} \ ,
\eeq
with $ {\hat T} = -\frac{V_0}{{\cal V}} {\hat M}^{-1} $.

\section{A Sum Rule for the Density of States}

The increment in the density of  states induced by the impurities is
\beq
\label{dens}
\delta \rho(\omega) = - \frac{1}{\pi} Im \sum_{\bk \nu} \delta G_{\bk
\bk }^{\nu \nu} (\omega + i 0^+) \ ,
\eeq
where
\beq
\begin{array}{cc}
&\delta G_{\bk \bk}^{\nu \nu} \equiv
-\frac{V_0}{{\cal V}} \
{\bf N}^T_{-\bk \nu} \cdot {\hat M}^{-1} \cdot {\bf N}_{\bk \nu} \ .
\end{array}
\eeq 
Recall that the summation over $\bk$ is restricted to the first
Brillouin zone. We can rewrite
\bea
\delta G_{\bk \bk}^{\nu \nu} & = & -\frac{V_0}{{\cal V}} \sum_{i,j} N^i_{-\bk \nu}
M^{-1}_{ij} N^j_{\bk \nu} \nonumber \\
 & = &  \sum_{i,j} M^{-1}_{ij} \left ( - \frac{V_0}{{\cal V}} N^j_{\bk \nu} N^i_{ -\bk
 \nu} \right ) \ .
\eea If we go to Matsubara frequencies and define
\[
G^0_{\bk \nu} = \frac{1}{ i \om - (-1)^\nu \omega_{\bk}} \ , \] we notice that
\[
\sum_{\bk \nu} \left  (- V_0 N^j_{\bk \nu} N^i_{ - \bk  \nu}  \right )
= \frac{\partial }{\partial i \om} {\hat M}_{ij} \ ,  \]
 and then \beq
\label{eqn1}
\sum_{\bk \nu} \delta G_{\bk \bk}^{\nu \nu} ( i \om ) = \frac{1}{{\cal V}} 
Tr \left [ {\hat M}^{-1} \frac{\partial {\hat M}}{\partial i \om} \right ] \ ,
\eeq
where the Trace is assumed to run over $\bk$, $\nu$ as well as over the
matrix indices. By convention we call
\beq
\delta G  \equiv \sum_{\bk \nu} \delta G_{\bk \bk}^{\nu \nu} \ .
\eeq The equation~(\ref{eqn1}) can also be written ($\ln Det = Tr \ln$)
\beq
\delta G ( i \om ) =\frac{1}{{\cal V}} \ \frac{\partial \left ( \ln Det 
{\hat M} \right ) }{\partial
i \om} \ .
\eeq 
We used this expression in a previous paper~\cite{cath} in order to
derive the additional density of states.
In this paper we will prefer to use the following expression
\beq
\delta G (i \om ) = 
\frac{1}{{\cal V}} 
\ Tr \left [ {\hat M}^{-2} \ \frac{\partial {\hat M}^2}{2 \ \partial
i \om} \right ] \ .
\eeq

\section{The matrix elements of ${\hat M}$}

As seen previously the evaluation of the density of states relies
on calculating the determinant of ${\hat M}$. We begin
by evaluating its matrix elements.
In this section, we just quote the result; the explicit calculation
being given in Appendix~\ref{ap2}.
In order to perform this calculation we make the assumption
that the energy $\omega \ll \Delta_0$. This will enable us to
linearize the spectrum for small energies.
\begin{figure}
\epsfxsize= 0.6 \textwidth
% *********** For one column  ********************
%\epsfxsize=7.0in 
% ***********************************8
\centerline{\epsfbox{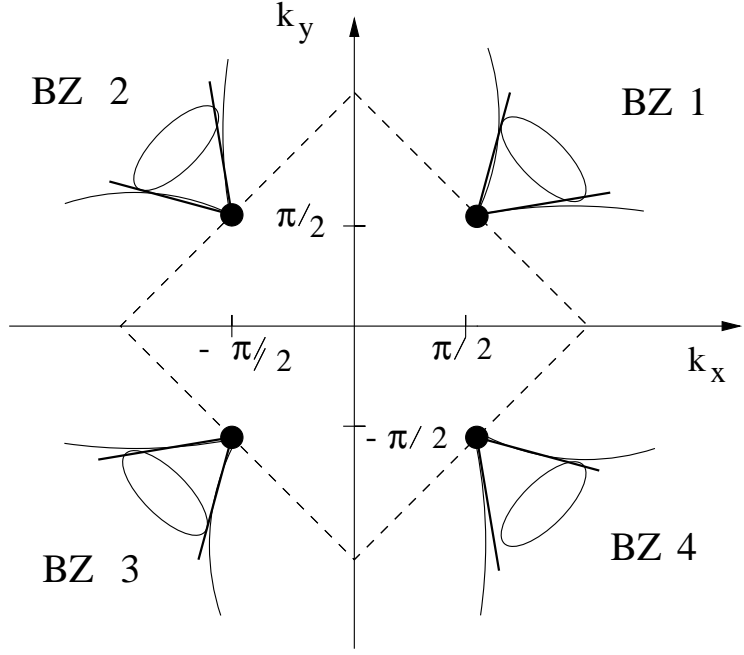}}
\vskip 0.1truein
\protect\caption{Linearization of the energy spectrum at very low
energy. In the first Brilloin zone we have four nodes centered at
$(\pm \pi/2, \pm \pi /2 )$ around which the energy spectrum is
linearized. There are some symmetry relations between the four
sud-zones called BZ 1, BZ 2, BZ 3 and BZ 4. The dotted line represents
the points where $\ek = \cos k_x + \cos k_y=0$. The linearized
spectrum around the nodes is represented in a third dimension in the
plane $(k_x, k_y )$. }
\label{BZ}
\end{figure}
Second we assume that $W=\Delta_0$ and the chemical potential $\mu =
0$ in order to simplify the calculation. We will see later what
happens when these two conditions are relaxed.
Under these conditions there are four nodes in the Brillouin zone
(cf. figure~\ref{BZ}) located at $ ( \pm \frac{\pi}{2} , \pm
\frac{\pi}{2} )$. With $\bk^\prime \equiv (\frac{\pi}{2},
\frac{\pi}{2} ) + \bk $ we get 
\bea
\omega_{\bk^\prime}^2 & = & W^2 \left [ \left ( \cos k^\prime_x + \cos
k^\prime_y -\mu \right )^2 + \left( \cos k^\prime_x - \cos k^\prime_y
\right )^2 \right] 
\nonumber \\
 & \simeq& 2 W^2 k^2 \ .
\eea 
Thus, $\omega_{\bk} =D k$ with $D = \sqrt{2} W$.
after integrating over the four nodes in the Brillouin zone, we find
\bea
A_{ij} & = & {\cal A}^0 (\bR_{ij}) + {\cal A}^1 (\bR_{ij}) \ , \nonumber
\\
C_{ij} & = & {\cal A}^0 (\bR_{ij}) - {\cal A}^1 (\bR_{ij}) \ , \nonumber \\
B_{ij} & = & {\cal B}^1 (\bR_{ij}) \ ,
\eea where
\bea
{\cal A }^0 (\bR) & \equiv & i \om \sum_\bk \frac{1}{ (i \om)^2 -\omega_{\bk}^2}
e^{-i \bk \cdot \bR} \nonumber \\
& = &  - {\cal F}^0 (\bR) \frac{i \om}{2 \pi D^2} K_0 \left ( \left |
\frac{R \omega_n}{D} \right | \right ) \ ,
\eea
whereby
\beq
\begin{array}{cc}
\label{pref1}
& {\cal F}^0 ( \bR) = 2 \cos \frac{\pi}{2}(R_x
+Ry) + 2 \cos \frac{\pi}{2} (R_x -R_y) \ ,
\end{array}
\eeq and $K_0$ is the Bessel function of rank zero. Note that
$|\omega_n| = \sqrt{- (i \om)^2}$. As shown in Appendix~\ref{ap2}, 
Eq.~(\ref{omn}),
\bea
{\cal A }^1 (\bR) & \equiv &  \sum_\bk \frac{\ek}{ (\om)^2 -\omega_{\bk}^2}
e^{-i \bk \cdot \bR}  \nonumber \\
& = &  {\cal F}^1 (\bR) \frac{\om}{2 \sqrt{2} \pi D^2} K_1 \left ( \left |
\frac{R \omega_n}{D} \right | \right ) \ , 
\eea
with
\beq
\begin{array}{cc}
\label{pref2}
& {\cal F}^1 ( \bR) = 2 \sin\frac{\pi}{2}(R_x
+Ry)( \cos \varphi + \sin \varphi ) + 2 \cos \frac{\pi}{2} (R_x -R_y)
(\cos \varphi -\sin \varphi ) \ ,
\end{array}
\eeq where $\varphi$ is the angle between $\bR$ and the x-axis and $K_1$ is
the Bessel function of rank one. In the same
manner 
\bea
{\cal B }^1 (\bR) & \equiv &  \sum_\bk \frac{\dk}{ (\om)^2 -\omega_{\bk}^2}
e^{-i \bk \cdot \bR} \nonumber \\
& = &  {\cal F}^2 (\bR) \frac{\om}{2 \sqrt{2} \pi D^2} K_1 \left ( \left |
\frac{R \omega_n}{D} \right | \right ) \ ,
\eea
with
\beq
\begin{array}{cc}
\label{pref3}
& {\cal F}^2 ( \bR) = 2 \sin\frac{\pi}{2}(R_x
+R_y)( \cos \varphi - \sin \varphi ) + 2 \cos \frac{\pi}{2} (R_x -R_y)
(\cos \varphi  + \sin \varphi ) \ .
\end{array}
\eeq
Note that the point ${\bf R} =0$ is rather special with $ {\cal A}^0
({\bf 0})=  \frac{4 \ i \om}{4 \pi D^2} \ln \left | \om / D \right |$
and with ${\cal A}^1 ({\bf 0}) ={\cal B}^1( {\bf 0}) =0$.

\section{Evaluation of the density of states}

\subsection{Unitary limit and low energies}
\label{seca}

In this section we will evaluate the leading term in the density of
states in the limit of low frequencies. When $0<R | \om | \ll D$
we have 
\beq
\begin{array}{cc}
{\displaystyle K_0 \left (\frac{R | \om |}{D} \right ) \simeq \ln
\left ( \frac{R |\om |}{D} \right )} \ ,
\; \; \; \; & {\displaystyle K_1 \left (\frac{R | \om |}{D} \right ) \simeq
\frac{D}{|\om| R} } \ ,
\end{array}
\eeq
so that in the limit $|\om | / D \rightarrow 0$ we have
\beq
\left | {\cal A}^0 ({\bf R}) \right | \ll \left | {\cal A}^1 ({\bf R}) \right | \ .
\eeq In the limit of low frequencies and for ${\bf R} \neq 0$, this
enables us to neglect ${\cal A}^0 ({\bf R})$ as compared to ${\cal
A}^1 ({\bf R})$ in the evaluation of the matrix ${\hat M}$ in Eq.~\ref{m}. 
For ${\bf R}={\bf 0}$, ${\cal A}^0 ({\bf 0})$ and ${\cal A}^1({\bf 0})$ are
negligible as compared to ${\cal A}^1( {\bf R} )$ for ${\bf R} \neq
{\bf 0}$. In the sequel we can safely avoid the point $ {\bf R} = {\bf 0}$ in
summations over ${\bf R}$ that occur during the evaluation of ${\hat M}^2$.
Second we notice that the  Bessel function $K_0$ and $K_1$ have an
exponential cut-off at $R_{max} = D / |\om | $ so that we can
safely use the approximation:
\beq
\label{eq43}
{\cal A}^1(\bR_{ij}) \simeq \left \{ \begin{array}{ll}
            {\displaystyle  \frac{ {\cal F}^1 (\bR_{ij}) }{2 \sqrt{2}
            \pi D R_{ij} } } \ , \; \; \; \;\mbox{if} & R < D
            /|\om | \ , \\
       0 \ , & \mbox{elsewhere} \ ,  \end{array} \right .
\label{eq: cal A}
\eeq
\beq
\label{eq44}
{\cal B}^1(\bR_{ij}) \simeq \left \{ \begin{array}{ll}
            {\displaystyle  \frac{ {\cal F}^2 (\bR_{ij}) }{2 \sqrt{2}
            \pi D R_{ij}} } \ , \; \; \; \;\mbox{if} & R < D
            /|\om | \ , \\
       0 \ , & \mbox{elsewhere} \ . \end{array} \right .
\label{eq: cal B}
\eeq
An important remark to make is that the $\omega$-dependence of
the matrix elements of ${\hat M}$ appears only through the upper
cut-off of the Bessel functions. 
In what follows, we make the crucial assumption of the unitary limit,
i.e., that $V_0 \rightarrow \infty$. Recalling the form of 
${\hat M}$ in equation~(\ref{m}) we see that in this limit the identity
matrix in ${\hat M}$ becomes negligible when compared to $A_{ij}$ and
$C_{ij}$.

\subsection{The divergences appear}

\begin{figure}
\epsfxsize= 0.6 \textwidth
% ***********For one column  ********************
%\epsfxsize=7.0in 
% ***********************************8
\centerline{\epsfbox{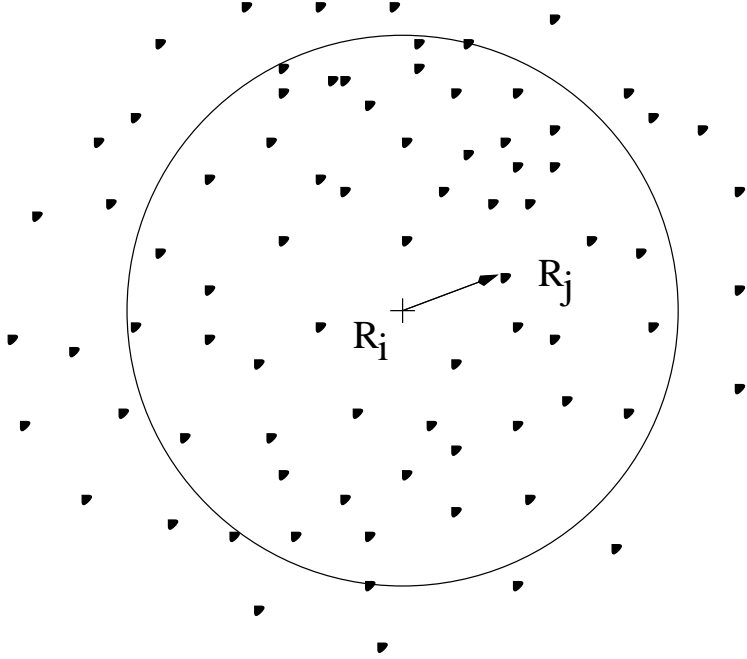}}
\vskip 0.1truein
\protect\caption{ The point $\bR_i$ and $\bR_j$ where the impurities
are located are represented in this figure. The integrand in
Eq.~(\ref{eqn10}) centered around the point $\bR_i$ is nonzero only
when $R_{ij} < D / |\om|$.}
\label{cont}
\end{figure}

Our aim is now to factorize the leading divergences in this problem. In order
to make the divergence apparent, it is more convenient to work with
\beq
\label{eq}
\delta G (i \om ) = Tr \left [ {\hat M}^{-2} \ \frac{\partial 
{\hat M}^2}{2 \ \partial
i \om} \right ] \ .
\eeq From section~\ref{seca}, we expect logarithmic factors $\ln \left
| D / \om \right |$ to appear. 
An important point to stress is that for any configuration of the
impurities, we will always find some factors $ \ln \left | D/ \om
\right | $ in ${\hat M}^2$. Within the unitary approximation we have
\beq
{\hat M}^2 = \left [ \begin{array}{cc}
{\hat A}^2 + {\hat B}^2 & {\hat A} {\hat B} - {\hat B} {\hat A} \\
{\hat A} {\hat B} - {\hat B} {\hat A} & {\hat A}^2 + {\hat B}^2
    \end{array} \right ] \ .
\eeq
To see that  $ \ln \left | D /
\om \right | $ is necessarily present in the diagonal terms
of ${\hat M}^2$, we estimate
\bea
\label{square}
M^2_{ii} & = & \sum_j \left ( A_{ij} A_{ji} + B_{ij} B_{ji} \right ) 
\nonumber \\
 & = & \sum_j \frac{1}{( 2 \pi D )^2}\frac{ \left[ ( {\cal F}^1 )^2 + ( {\cal
 F}^2 )^2 \right]}{ R_{ij}^2} \ ,
\eea
where the summation over $j$ is restricted to $ 0< R_{ij} < D / | \om | $.
Provided the impurities are rather homogeneously scattered in the
system (around each impurity site $\bR_i$ one can find a macroscopic
amount of impurities inside a circle of radius 
$ D/ | \om | $, cf. figure~\ref{cont} ), we can
take the continuous limit,
\beq
\label{eqn10}
{\hat M}^2_{i i} \sim \frac{ 2 \pi V_0}{(2 \pi D )^2} \int_1^{D/| \om|} \frac{ ({\cal
F}^1) ^2 + ( {\cal F}^2)^2 }{ R} \ dR \ .
\eeq Now $ ({\cal
F}^1) ^2 + ( {\cal F}^2)^2$ are oscillatory but always positive so for
all impurity at site $\bR_i$ we have
\beq
\label{mii}
{\hat M}^2_{i i} = C \ln \left | \frac{D}{\om} \right | \ ,
\eeq where $C$ is a constant.

The biggest coefficients in ${\hat M}^2$ are situated on the
diagonal. To see it, we distinguish between the off-diagonal elements
that are diagonal in the particle-hole grading and  those that are not.
The magnitude of off-diagonal elements
that are diagonal in the particle-hole grading can be written
\beq
{\hat M}^2_{i k} = \sum_j \frac{1}{2 \pi D} \frac{ \left [ {\cal F}^1_{ij}{\cal F}^1_{jk} + {\cal F}^2_{ij}{\cal F}^2_{jk} \right ]}{R_{ij}R_{jk}} \ .
\eeq As $ i \neq k$, ${\hat M}^2_{ik}$ picks up an oscillatory prefactor (a combination of $ e^{\pm i \pi /2 R_{ik}} e^{\pm i \varphi}$ as in eqns.~ (\ref{pref1}), (\ref{pref2}), (\ref{pref3}) ). Furthermore, the logarithmic divergence gets rescaled by $R_{ik}$:
\beq
{\hat M}^2_{ik} = C \ Osc({\bf R}_{ik})\ln \left | \frac{D R_{i k}}{\om} \right | \ ,
\eeq where $\left | Osc ({\bf R}_{ik}) \right | \leq 1$.
Hence for all sites $k \neq i$ we have
\beq \left | \frac{{\hat M}_{ik}^2}{{\hat M}_{ii}^2} \right | \leq 1 \ .
\eeq

\begin{figure}
\epsfxsize= 0.6 \textwidth
% ***********For one column  ********************
%\epsfxsize=7.0in 
% ***********************************8
\centerline{\epsfbox{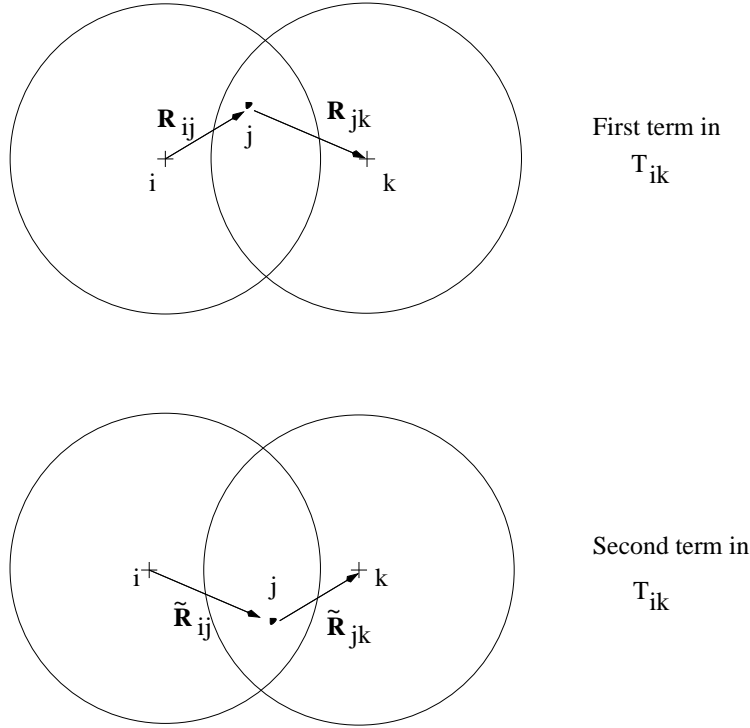}}
\vskip 0.1truein
\protect\caption{Vanishing of the element $T_{ik}$. 
The summation zone is the surface of intersection of the two
disks. The first term in~(\ref{tik}) is represented in the upper
drawing whereas the second one is represented in the lower drawing. For each
summation point $\bR_j$ in the upper drawing there is a symmetric one
${\tilde \bR}_j$ 
in the lower drawing such that $ \bR_{ij} = {\tilde \bR}_{jk}$ and
${\tilde \bR}_{ij} = \bR_{jk}$.}
\label{figtik}
\end{figure}

In turn,
the terms in the off-diagonal blocks of ${\hat M}^2$ with respect to the 
particle-hole vanish. Indeed, 
if we denote the off-diagonal elements of $\hat M^2$
with respect to the particle-hole grading by
\beq
T_{i k} \equiv A_{i j} B_{j k } - B_{i j} A_{j k}\ ,
\eeq 
we have
\beq
\label{tik}
T_{ik} = \sum_j \frac{1}{( 2 \pi D)^2} \left ( \frac{ {\cal F}^1
(\bR_{i j} ) {\cal F}^2 (\bR_{j k} ) }{ R_{i j} R_{j k} } - \frac{ {\cal F}^2
(\bR_{i j} ) {\cal F}^1 (\bR_{j k} ) }{ R_{i j} R_{j k} } \right ) \ ,
\eeq
where here the sum runs over the points $ \bR_j $ such that $ R_{i j} <
D/ |\om | $ {\it and} $ R_{ j k} < | \om | $.
Noticing the symmetry of ${\cal F}^1$ and $ {\cal F}^2 $ under the
transformation $\varphi \rightarrow \varphi + \pi $, we can show that
the two terms on the r.~h.~s.~of Eq.~(\ref{tik}) cancel identically. 
Indeed the integrands
$\frac{{\cal F}^1(R_{ij}) }{R_{i j}}$ and $\frac{{\cal F}^2 (R_{ik}
)}{R_{i k}}$  are represented respectively within each circle of
figure~\ref{figtik} 
( both of these terms have a cut-off at $ R_{max} = D/| \om
| $). The summation zone is the surface of intersection of the two
disks. The first term in~(\ref{tik}) is represented in the upper
drawing whereas the second one is represented in the lower drawing. For each
summation point $\bR_j$ in the upper drawing there is a symmetric one
${\tilde \bR}_j$ 
in the lower drawing such that $ \bR_{ij} = {\tilde \bR}_{jk}$ and
${\tilde \bR}_{ij} = \bR_{jk}$.
Thus $T_{ik} = 0$.

In conclusion we find it convenient to define
a matrix ${\hat S}$ by
\beq
\label{m2}
{\hat M}^2 = C \ln \left | \frac{D}{
\om} \right | \ {\hat S} \ ,
\eeq where $C$ is the constant defined in (\ref{mii}), independent of disorder. 
The matrix ${\hat S}$ depends on the particular
configuration of the impurities in the system. It satisfies
\beq
\label{bound}
\left | S_{i j} \right | \leq  1 \ .
\eeq

\subsection{ Asymptotic value of the density of states}

Substituting the value of ${\hat M}^2$ from~(\ref{m2}) into the
equation~(\ref{eq}) we  find
\bea
\label{eq100}
\delta G (i \om ) & = &  
{\cal T}_{div} ( i \om ) 
+
{\cal R} ( i \om ) ,
\eea
where $\la \cdot \ra$ denotes the average over disorder,
${\hat I}$ is the $2 N \times 2 N $ density matrix,
and
\beq
\begin{array}{cc}
{\cal T}_{div} ( i \om ) \equiv {\displaystyle \frac{1}{2 {\cal V}} \ 
 \frac{\partial}{\partial i \om} \ln  \ln \left ( D /
| \om | \right )} \ Tr \  {\hat I}\ , \; \; \; \; 
\; \; \; \; 
{|cal R} ( i \om )  \equiv 
{\displaystyle \frac{1}{2 {\cal V}} \la Tr {\hat S}^{-1} \frac{\partial
}{\partial i \om } {\hat S} \ra } \ . \end{array} 
\eeq
Then the first term in Eq.~(\ref{eq100}) ${\cal T}_{div}$ is responsible for the
singular density of states that we obtain. Indeed it gives rise to
\beq
{\cal T}_{div} (i \om) = \frac{N}{{\cal V}} \frac{1}{ i \om \ \ln | \om / D
|} \ ,
\eeq
where we recall that $N$ is the number of impurities.
After analytic continuation  (remember that $| \om|= \sqrt{- (i
\om)^2}$, Eq.~(\ref{omn}) ) and assuming that the reminder ${\cal R}$ in
Eq.~(\ref{eq100}) is negligible, we get
\beq
\delta \rho (\omega) \simeq - \frac{1}{\pi} n_i Im \left [ \frac{1}{
(\omega + i \delta ) \ \ln \left ( \frac{i \omega + \delta}{D} \right
) } \right] \ ,
\eeq
and thus
\beq
\label{result}
\label{res}
\begin{array}{ll}
& \delta \rho (\omega ) \simeq {\displaystyle \frac{n_i}{2}
\frac{1}{ | \omega | \left [ \ln^2 ( | \omega | /  D ) + ( \pi /2)^2
\right ] } } \ ,
\end{array}
\eeq
where $n_i$ is the density of impurities in the system.
We note that this expression is normalizable:
\beq
\int_{-D}^{D} \delta \rho (\omega) d \omega = 2 n_i \  .
\eeq
In order to prove the result~(\ref{result}) we still have to show that the 
reminder ${\cal R}$
in~(\ref{eq}) is negligible as compared to ${\cal T}_{div}$. In order to do
this we have to give some insight about the form of ${\hat S}^{-1}$.

\subsection{The form of ${\hat S}^{-1}$}

\begin{figure}
\epsfxsize= 0.6 \textwidth
% ***********For one column  ********************
%\epsfxsize=7.0in 
% ***********************************8
\centerline{\epsfbox{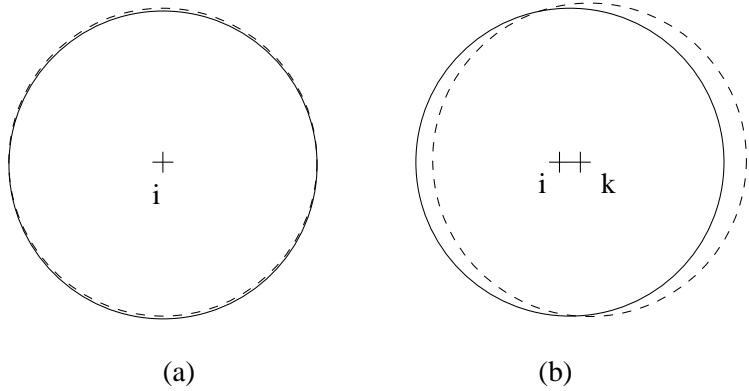}}
\vskip 0.1truein
\protect\caption{Volume of summation of $\sum_j {\hat S}_{ij} 
{\hat S}^{-1}_{jk}$ in two configurations. In $(a)$ we have $ i = k$ and in
$(b)$ we have $i \neq k $ but $i$ still very close to $k$.}
\label{figinv}
\end{figure}

The matrix ${\hat S}$ is invertible (since ${\hat M}$ is invertible) and we will find a reasonable
candidate to the inverse of ${\hat S}$ in order to give an estimation
of the reminder ${\cal R}$. 
Inverting ${\hat S }$ means 
we can find a matrix ${\hat S}^{-1}$ such that for any given pair of sites
$i$ and $k$ 
\beq
\label{inv}
\sum_j {\hat S}_{i j} {\hat S}^{-1}_{j k} = \delta_{i k} \ .
\eeq
We introduce a pictorial representation of ${\hat S}$ by drawing a disk ( called ${\hat S}$-disk) of radius $ |D / \om|$ centered at $\bR_i$. To each location $\bR_j$ of an impurity there corresponds
a matrix element $S_{ij}$ which depends on 
the vector $\bR_{i j}=\bR_i-\bR_j$. 
From Eqs.~(\ref{eq: cal A},\ref{eq: cal B}) we recall that
\beq
\left \{ \begin{array}{lcl}
      |S_{i j} | \leq 1 \ , \; \; \; &  \mbox{ for} & \; \; R_{i j} < | D /
      \om | \ , \\
    S_{i j} =0 \ , \; \; \; & \mbox{for} & \; \; R_{i j} \geq | D /
      \om | \ ,
    \end{array} \right  .
\eeq
inside the disk, ${\hat S}$ has some non vanishing matrix elements
$ | S_{i j} | < 1 $. Outside the disk, $ S_{i j} = 0 $.
It's important to note at this point that the only dependence on
$\omega$ in the ${\hat S}$ matrix comes from the cut-off.
In order to satisfy~(\ref{inv}) we must presume that ${\hat S}^{-1}$
has the same cut-off $ | D / \om |$ as ${\hat S}$.
Hence we represent again ${\hat S}^{-1}_{i k}$ by a disk (called ${\hat S}^{-1}$-disk), but centered
around $\bR_k$ this time. Now the summation $ \sum_j {\hat S}_{i j}
{\hat S}^{-1}_{j k} $ runs over the intersection of the ${\hat S}$-disk and the ${\hat S}^{-1}$-disk.
The key difficulty in order to invert ${\hat S}$ is to find a matrix ${\hat S}^{-1}$,
such that when the two disks have the same center we have
\[ \sum_j {\hat S}_{i j} {\hat S}^{-1}_{j i} =1 \ , \]
whereas when the two centers differ, even by a small amount, we have 
\[
\begin{array}{ll}
\sum_j {\hat S}_{i j} {\hat S}^{-1}_{j k} = 0 \ ,  \; \; \; \; & k \neq
i\ . \end{array} \]
This is illustrated on figure~\ref{figinv} where on the left side (case (a)) the
two circles are centered at the same point and on the right side (case (b)) the
two centers differ by a tiny amount. In both cases the intersecting area of the two disks is almost identical, but in case (a) the
result has to be $1$ whereas in case (b) it has to be $0$.

The matrix elements of ${\hat S}$ inside the disk are random, but the condition $| {\hat S}_{i j}  | \leq 1$ is independent of the realization of disorder. The worst possible
situation for differentiating between cases (a) and (b) is
when all the matrix elements inside the ${\hat S}$-disk
have their maximum value $1$.

We define the {\it external boundary} of the ${\hat S}^{-1}$-disk. By {\it external boundary} we mean the circle exactly adjacent to the disk and external to it.
Upon the external boundary the matrix elements of ${\hat S}^{-1}$ are defined as non vanishing and negative, so that the summation over it compensates the summation over the intersection of the ${\hat S}$-disk and the ${\hat S}^{-1}$-disk.
In case (a) the external boundary doesn't
touch the ${\hat S}$-disk and thus has no effect upon the summation over the intersection of the ${\hat S}$-disk and the ${\hat S}^{-1}$-disk.
Alternatively in case (b) the summations over
the intersection of the ${\hat S}$-disk and the ${\hat S}^{-1}$-disk and over the external boundary cancel out.

Let's take an explicit example where
$S_{i j}= 1$ if $R_{i j} < | D / \om |$ and $S_{i j} = 0$ elsewhere.
We define ${\hat S}^{-1}_{i j} $ in the following way. For  $R_{ij} <
| D / \om |$, ${\hat S}^{-1}_{i j} $ is proportional to a random
configuration of $\pm 1$ such that after integration over a disk of
volume $V= \pi \left | D / \om \right |^2 $ we get 
\bea
\sum_{V} \pm 1 & = &  \sqrt{V}  \nonumber \\
  & = & \sqrt{\pi} \frac{D}{| \om |} \ .
\eea
By the central limit theorem, there are many random configurations of
$\pm 1$ that verify this condition.
We then take the external boundary to be proportional to $(- 1 /
\sqrt{\pi} )$ with the same proportionality constant.
Thus 
\beq
\left \{ \begin{array}{ll}
{\hat S}^{-1}_{ij} = A ( \pm 1) \ , \; \; \; \; &\mbox{if} \; \; R_{ij} <
| D / \om | \ , \\
{\hat S}^{-1}_{ij} =  
- A {\displaystyle  \frac{1}{\sqrt{\pi}} }\ , \; \; \; \;& \mbox{if}
\; \; R_{ij} = | D / \om | \ , \\
{\hat S}^{-1}_{ij} = 0 \ ,  & \mbox{if} \; \; R_{ij} >
| D / \om | \ ,
\end{array} \right .
\eeq
where $A$ is a constant.
We notice that when $\bR_i$ and $\bR_k$ are infinitely close, then 
\[
\sum_{boundary} {\hat S}^{-1}_{ij} = - A \sqrt{\pi} \frac{D}{| \om |}
\ ,
\]
and exactly compensates the summation over the volume inside.
The proportionality constant $A$ is fixed so that 
\[
\sum_j {\hat S}_{ij} {\hat S}^{-1}_{ij} =1 \ , \] 
\[
\begin{array}{ll}
\mbox{thus} \; \; \; \; & A = {\displaystyle \frac { | \om |
}{\sqrt{\pi} D} } \ .
\end{array}  \]

In summary
\beq
 \left \{ \begin{array}{ll}
{\hat S}^{-1}_{ij} = {\displaystyle \frac { | \om | }{\sqrt{\pi} D} }
( \pm 1) \ , \; \; \; \; &\mbox{if} \; \; R_{ij} <
| D / \om | \ , \\
{\hat S}^{-1}_{ij} =  - {\displaystyle \frac { | \om | }{\pi D} } \ , \; \; \; \; & \mbox{if}
\; \; R_{ij} = | D / \om | \ , \\
{\hat S}^{-1}_{ij} = 0 \ , & \mbox{if} \; \; R_{ij} >
| D / \om | \ .
\end{array} \right .
\eeq

\begin{figure}[here]
\epsfxsize= 0.5 \textwidth
% ***********For one column  ********************
%\epsfxsize=7.0in 
% ***********************************8
\centerline{\epsfbox{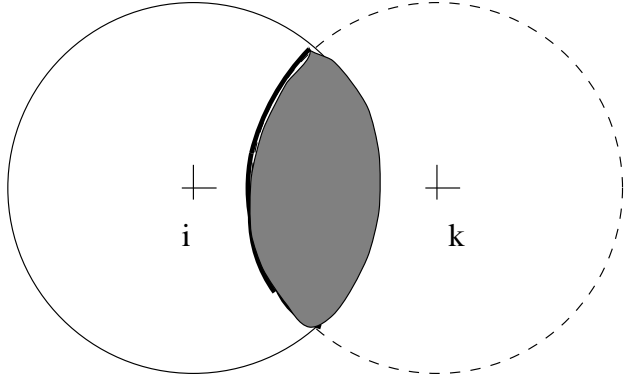}}
\vskip 0.1truein
\protect\caption{Compensation between the volume and boundary.}
\label{figint}
\end{figure}

\begin{figure}[here]
\epsfxsize= 0.5 \textwidth
% ***********For one column  ********************
%\epsfxsize=7.0in 
% ***********************************8
\centerline{\epsfbox{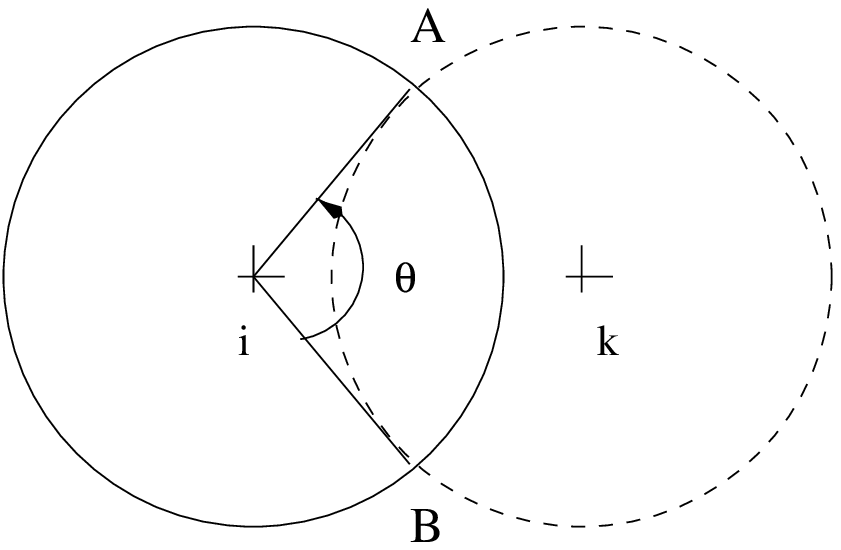}}
\vskip 0.1truein
\label{figint2}
\end{figure}

Now for each intermediate case where $k \neq i$
(cf. figure~\ref{figint}) we want to be sure that the intersection of the ${\hat S}$-disk and the ${\hat S}^{-1}$-disk compensates the sum over the external boundary of ${\hat
S}^{-1}$ which crosses the ${\hat S}$-disk. This is obviously not
the case for any configuration of random $\pm 1$ in ${\hat S}^{-1}$
but we believe there is one (and actually only one because there is only
one inverse for ${\hat S}$!) configuration
which satisfies it for all positions of $\bR_i$ and $\bR_k$.
As represented in figure~\ref{figint2} we call $\theta$ the angle
made by the center $i$ with the two points $A$ and $B$, intersection between the two circles. The
area of summation is given by $A_\theta = \theta R^2 - 4 R^2 \sin
\theta$ and the intersecting arc's length by $L_\theta = \theta R$.
They both scale in respectively $R^2$ and $R$ with varying prefactors.
So the desired configuration of random $\pm 1$ in the ${\hat S}^{-1}$-disk has to be ``denser''
towards the center of the circle than towards the boundary.
We are not able to write down explicitly this configuration of random
$\pm 1$ inside the area of the ${\hat S}^{-1}$-disk, but actually it doesn't
matter, because as we will see in the next paragraph the evaluation of
the reminder doesn't depend on it.

\subsection{Evaluation of the reminder $ {\cal R}$}

\begin{figure}
\epsfxsize= 0.5 \textwidth
% ***********For one column  ********************
%\epsfxsize=7.0in 
% ***********************************8
\centerline{\epsfbox{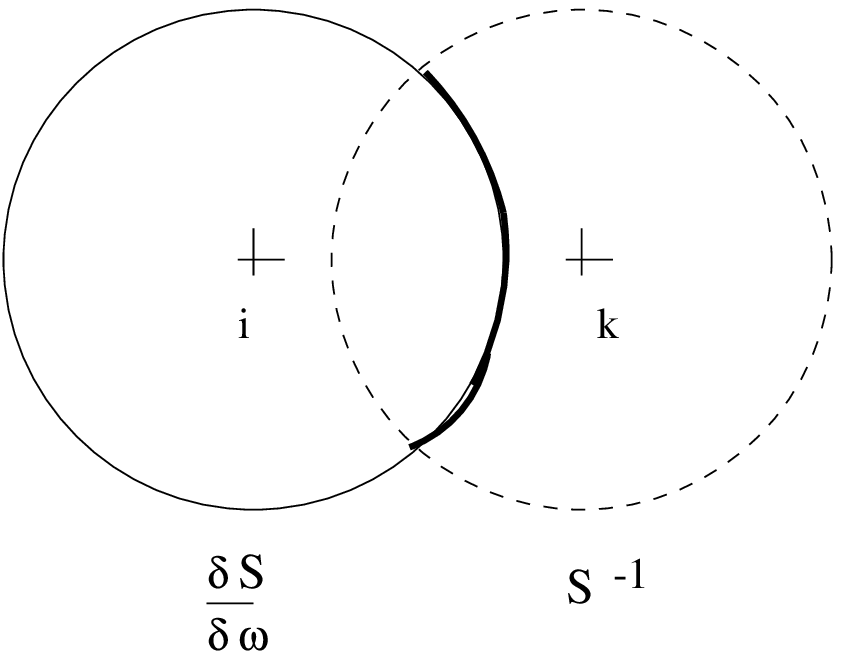}}
\vskip 0.1truein
\protect \caption{Summation over $j$ in ${\displaystyle \frac{\partial {\hat S}}{\partial \om}}$.}
\label{figfin}
\end{figure}

Now we evaluate
\beq
{\cal R} = {\displaystyle \frac{1}{2 {\cal V}} \la Tr {\hat S}^{-1} \frac{\partial
}{\partial i \om } {\hat S} \ra } \ .
\eeq
An important point is that as ${\hat S}$ depends on $ \om $ only via
its boundary, $\partial {\hat S} / \partial \om $ is a matrix with non
zero values only on the external boundary of the ${\hat S}$-disk.
explicitly, taking our example where ${\hat S}_{ij} =1$ if $R_{ij} < D
/ | \om | $ we get
\beq
\left \{ \begin{array}{ll}
       -{\displaystyle \frac{\partial {\hat S}}{\partial \om} =1 } \ , \; \; \; \; &
       \mbox{if} \; \; R_{ij} = D/ | \om | \ , \\
{\displaystyle \frac{\partial {\hat S}}{\partial \om} =0 } \ , &
       \mbox{elsewhere} \ .
\end{array} \right .
\eeq
Since the matrix elements in the ${\hat S}^{-1}$-disk have random (positive and negative) signs 
whereas upon the external boundary they have constant sign (negative here),
the maximum value of $\left | {\hat S}^{-1} \partial {\hat S} /
\partial \om \right |$ is reached when $i = k$, that is when the external boundary of the ${\hat S}$-disk 
(where the matrix elements of $\partial {\hat S} / \partial \om$ are non vanishing) matches the external boundary of the
${\hat S}^{-1}$-disk (cf. Fig.~\ref{figfin}).

In our special case we get
\bea
\left | {\hat S}^{-1} \frac{\partial {\hat S}}{\partial \om } \right |
& =  & \left ( \frac{ | \om | }{ \pi D} \right ) \left ( \frac{ 2 \pi
D}{ | \om |} \right) \ , \nonumber \\
 & = & 2 \ .
\eea
Since $\left | \partial {\hat S} / \partial \om
\right |$ has its maximum value when all the matrix
elements of ${\hat S}$ inside ${\hat S}$-disk equal $1$, we
have indeed evaluated an upper bound for the reminder ${\cal R}$.
The reminder ${\cal R}$ is thus negligible compared to the leading divergence in the density of states
in the limit $| \om | / D \ll 1$.

\section{Discussion}

The first question to address is what happens when the different
conditions under which our calculation was performed are relaxed.
First consider the more realistic situation where the bandwidth $W$
and the superconducting gap $\Delta_0$ are not equal (experimentally,
we have $\Delta_0 \simeq 0.10 \ W$). According to 
Ref.~\onlinecite{repbalat} the
power law dependence of the matrix elements of ${\hat M}$
(cf. equations~(\ref{eq43}) and~(\ref{eq44})) is still preserved, but gets an overall
prefactor of $\Delta_0 / W$. The scale under which our calculation is
valid is now $ |\omega| \ll V_0 \Delta_0 / W$. In addition, the hopping matrix ${\hat M}$ will
show a strong spatial anisotropy~\cite{balatsky}. This
anisotropy can be absorbed into the overall prefactors $ {\cal F}^0$,
${\cal F}^1$ and ${\cal F}^2$ (resp. eqns.~(\ref{pref1}),
(\ref{pref2}) and (\ref{pref3})) entering the definition of the matrix
elements $A_{ij}$, $C_{ij}$ and $B_{ij}$. Since only the square of
these factors enter the leading divergence (cf. Eq.~(\ref{square})), we believe the result
stays unchanged.

What happens if the bandwidth are not symmetric anymore, that is if
$\mu \neq 0$ but still $\mu \ll \Delta_0$ ?

First the nodes are moved away from the point $ \left ( \pm \pi/2, \pm
\pi/2 \right )$ so that transversal nodes are now separated by the
vectors $ {\bf Q}= \left ( \pi (1-\delta), \pi (1-\delta) \right )$
and ${\bf Q}^*= \left (- \pi (1-\delta), \pi (1-\delta) \right )$
where $ \delta= \mu / \Delta_0 $ and $\mu$ is the increase in the chemical
potential. This leads to a change of the phase factors in $A_{ij}$ and
$B_{ij}$. Namely we get $ {\cal F}^0 (R) =2 \left [ \cos \left ( {\bf Q} \cdot
{\bf R} /2 \right ) + \cos \left ( {\bf Q}^* \cdot {\bf R} /2 \right ) \right ]$; 
\bea
{\cal F}^1 (R)&  = & 2 \left [ \sin \left ( {\bf Q} \cdot {\bf R} / 2
\right ) \left ( \cos \varphi + \sin \varphi \right ) \right . \ , \nonumber \\
  &  - & \left . \sin \left ( {\bf Q}^* \cdot {\bf R} / 2 \right ) \left ( \cos \varphi - \sin \varphi \right ) \right ] \nonumber \ ,
\eea and

\bea
{\cal F}^2 (R) & = &2 \left [ \sin \left ( {\bf Q} \cdot {\bf R} / 2
\right ) \left ( \cos \varphi - \sin \varphi \right )  \right . \ , \nonumber \\
 & - & \left . \sin \left ( {\bf Q}^* \cdot {\bf R} / 2 \right ) \left ( \cos \varphi + \sin \varphi \right ) \right ] \nonumber \ .
\eea
As $\delta$ is a small parameter, this change in the phase won't affect the existence of the logarithmic divergence in ${\hat M}^2$.
Additionally, away from half filling, the bands of quasi particles
and quasi holes become asymmetric to account for the removing of
particles in the system. The difference induced in $ A_{ij}$,
$B_{ij}$ and $C_{ij}$ comes from the highest part of the energy
spectrum, where $k \simeq 1$ and $\omega \simeq D$ and shouldn't
affect our result.

Our solution is valid under the assumption of unitary limit, meaning
that $V_0$ is the largest scale in the problem. As shown in~\cite{autres1,khaliullin,balatsky}, 
to non interacting single impurities is associated  the creation of a bound states decaying as $1 / (R \ln R)$.
Following Balatsky {\it et
al.}~(\cite{balatsky}) this would lead to delocalization due to the
formation of an impurity band. The result we find for the
density of states is indeed reminiscent
of a Dyson-like singularity ( in d=1, a Dyson-like singularity
corresponds to a density of state diverging in $1/\left ( |\omega|
\ln^3(|\omega| /D) \right )$ ), associated in one dimension with delocalization~\cite{thouless}. 

What happens when the condition of unitarity is relaxed is still very
much an open problem. In the case of weak disorder, some $\sigma$-model analysis have
been performed~\cite{alexei,mudry,senthil} concluding to a vanishing density
of state under an energy scale $E_2= D_F / \xi^2$, where $D_F$ is the bare
diffusion constant and $\xi$ is the localization length. This result
is strongly supported by symmetry considerations. Indeed, a disordered
d-wave superconductor belongs to class C$_{I}$, according to the
classification of ref.~\cite{atland}, meaning that the Hamiltonian is
invariant under time-reversal symmetry as well as spin  rotation
symmetry. According to random matrix theory a universal behavior is
expected, inducing a vanishing density of states on the scale of the
level spacing induced by the finite localazation length.

Consider one impurity with scattering potential $V_0$. The effective
potential at the impurity site can be evaluated
exactly~\cite{autres1,khaliullin} and is given by $ {\bar V} =
\frac{V_0}{1 - V_0 |\omega| \ln \left | D / \omega \right | }$. In the
unitary limit ( $ V_0 \rightarrow \infty$) we get $ {\bar V} =
\frac{-1}{|\omega | \ln \left | D / \omega \right |}$. This effective
potential diverges when $\omega $ goes to zero.
On the other hand, we notice that the derivation of non linear sigma
models for disordered systems requires Gaussian disorder, and
especially require that the average effective disorder potential
vanishes $ \langle V \rangle = 0 $ (the second moment is nonzero). If
$ \langle V \rangle$ is nonzero but is a constant as a function of
energy, it can be absorbed as a redefinition of the chemical
potential, but the case where the effective potential would diverge as
$\omega$ goes to zero belongs to another universality class: the energy of an effective non-linear sigma model
would renormalized to $\omega - \Sigma (\omega)$, where $\Sigma (\omega)$ diverges when $omega$ goes to zero.
The energy scale under which the non-linear sigma model describes the diffusive modes then becomes inaccessible 
since the effective energy never gets close to zero.
In our problem the result obtained on the density of states indicates that the self-energy is of the form $\Sigma(\omega) = \frac{1}{ \left | \omega \ln^2 (\omega / D) \right |}$, diverging as $\omega$ goes to zero, but still different from the one impurity case.
The feynmann diagrams leading to such this self-energy will be studied in a future publication~\cite{preparation}.
We believe the method presented here, using the T-matrix equation
takes care in a non perturbative way of the leading divergence in the unitary limit.
One possible scenario
which would reconciliate the two limits of weak and strong disorder is
that the unitarity limit fixed point is unstable (as soon as $V_0$
becomes finite, the effective potential saturates), but the
cross-over regime close to unitarity is still very much influenced by
the strong disorder fixed point.

\vskip 1 cm
We would like to thank A.V. Balatsky, P. Coleman, M. Hettler,
R. Joynt, C. Mudry, R. Narayanan, A. M. Tsvelik, T. Xiang for useful discussions related to
this work. We are especially grateful to J. Chalker and B.D. Simon for
discussions concerning symmetries of disordered d-wave superconductors. As
this paper was submitted, a recent numerical study~\cite{num} confirmed the singular density of states
in the unitary limit. However they concluded that the resonant density of states alone was not
sufficient to induce delocalisation.

This work is supported by NSF Grant No. DMR 9813764 and by (CP) a
Bourse Lavoisier and the research fund from the EPSRC, UK.

\pagebreak

\appendix

\section{Derivation of the T-matrix Equation }
\label{ap1}

Starting form equation~(\ref{mo}), replacing the lagrangian
of~(\ref{lag}) into it and multiplicating from the left by $G^0$ gives
\bea
\label{int}
G^{\nu \nu^\prime}_{\bk \bk^\prime} & + & \frac{V_0}{{\cal V}} 
\sum_i e^{i \bk
\cdot \bR_i} (-1)^\nu t_{\bk \nu} G^0_{\bk \nu} G^1_{\bk^\prime \nu^\prime}
(\bR_i) \nonumber \\
&+& \frac{V_0}{{\cal V}} \sum_i e^{i \bk \cdot \bR_i} t_{\bk \nu+1}
G^0_{\bk \nu} G^2_{\bk^\prime \nu^\prime} = G^0_{\bk \nu} \delta_{\bk
\bk^\prime} \delta_{\nu \nu^\prime} \ ,
\eea
where
\beq
\begin{array}{l}
G^1_{\bk \nu}(\bR_i) \equiv - \sum_{\bq m} e^{-i \bq \cdot \bR_i} (-1)^m
t_{\bq m} G^{m \nu}_{\bq \bk} \ , \\
G^2_{\bk \nu}(\bR_i) \equiv \sum_{\bq m} e^{-i \bq \cdot \bR_i} t_{\bq
m+1} G_{\bq \bk}^{m \nu} \ .  \end{array}
\eeq
Now we have two unknown functions $G^1$ and $G^2$ that we evaluate
using formula~(\ref{int}):
\bea
-G^1_{\bk \nu}(\bR_j) & + & \frac{V_0}{{\cal V}} \sum_i \sum_{\bq m} e^{i
\bq \cdot ( \bR_i -\bR_j)} (t_{\bq m})^2 G^0_{\bq m} G^1_{\bk
\nu}(\bR_i) \nonumber \\
& + & \frac{V_0}{{\cal V}} \sum_i \sum_{\bq m} e^{i \bq \cdot (\bR_i +
\bR_j)} (-1)^m t_{\bq m} t_{\bq m+1} G^0_{\bq m} G^2_{\bk \nu}(\bR_i) =
(-1)^\nu e^{-i \bk \cdot \bR_j} t_{\bk \nu} G^0_{\bk \nu} \ ;
\eea 
\bea
G^2_{\bk \nu}(\bR_j) & + & \frac{V_0}{{\cal V}} \sum_i \sum_{\bq m} e^{i
\bq \cdot ( \bR_i -\bR_j)} (-1)^m t_{\bq m} t_{\bq m+1} G^0_{\bq m} G^1_{\bk
\nu}(\bR_i) \nonumber \\
& + & \frac{V_0}{{\cal V}} \sum_i \sum_{\bq m} e^{i \bq \cdot (\bR_i +
\bR_j)} (t_{\bq m})^2 G^0_{\bq m} G^2_{\bk \nu}(\bR_i) =
 e^{-i \bk \cdot \bR_j} t_{\bk \nu+1} G^0_{\bk \nu} \ .
\eea These two equations can be rewritten matricially as
\beq
\label{eq322}
\begin{array}{l}
\left ( - \delta_{ij} + V_0 A_{ij} \right ) G^1_{\bk \nu}(\bR_j) + V_0
B_{ij} G^2_{\bk \nu} (\bR_j) =N^1_{\bk \nu}(\bR_j) \ , \\
\left ( \delta_{ij} + V_0 C_{ij} \right ) G^2_{\bk \nu}(\bR_j) + V_0
B_{ij} G^1_{\bk \nu}(\bR_j) = N^2_{\bk \nu}(\bR_j) \ .
\end{array}
\eeq
Define the $2 N$ vector ${\bf V}$
\beq
\begin{array}{ccc}
{\bf V} \equiv \left ( \begin{array}{c} {\bf V}^1 \\ {\bf V}^2 \end{array}
\right ) \ , \; \; \; & {\bf V}^1_{\bk \nu} \equiv \left ( \begin{array}{c} 
G^1_{\bk \nu}(\bR_1) \\ \vdots \\ G^1_{\bk \nu}(\bR_N) \end{array} \right
) \ , \; \; \; & {\bf V}^2_{\bk \nu} \equiv \left ( \begin{array}{c} 
G^2_{\bk \nu}(\bR_1) \\ \vdots \\ G^2_{\bk \nu}(\bR_N) \end{array} \right
) \ ,
\end{array}
\eeq
and the equation~(\ref{eq322}) can be written
\beq
\label{eqa7}
{\hat M} {\bf V}_{\bk \nu} = {\bf N}_{\bk \nu} \ .
\eeq
But then equation~(\ref{int}) becomes
\beq
G_{\bk \bk^\prime}^{\nu \nu^\prime} + \frac{V_0}{{\cal V}} {\bf N}_{\bk \nu} \cdot {\bf
V}_{\bk \nu} = G^0_{\bk \nu} \delta_{\bk \bk^\prime} \delta_{\nu \nu^\prime} \
.
\eeq
 Insertion of~(\ref{eqa7}) yields
\beq
G_{\bk \bk^\prime}^{\nu \nu^\prime} = G^0_{\bk \bk^\prime} -
\frac{V_0}{{\cal V}} \; \ {\bf N}^T_{- \bk  \nu} \cdot {\hat M}^{-1} \cdot {\bf
N}_{\bk^\prime \nu^\prime} \ .
\eeq

\section{Calculation of the matrix elements $A_{ij}$, $B_{ij}$ and
$C_{ij}$}
\label{ap2}
\bea
A_{ij} &  = & \sum_n \int \frac{d^2 k}{(2 \pi)^2} 
e^{- i \bk \cdot (\bR_i -\bR_j)}
(t_{\bk n} )^2 G^0_{\bk n}  \\
  & = & \int \frac{d^2 k}{(2 \pi)^2} e^{-i \bk \cdot \bR_{ij} } \left (
\frac{ u_\bk^2}{ i \om - \omega_{\bk}} + \frac{v_\bk^2}{i \om + \omega_{\bk}}
\right ) \ ,
\eea 
where the integration runs over the first Brillouin zone. Since
\beq
\begin{array}{l}
 u_\bk^2 = \frac{1}{2} \left ( 1 + \frac{ \ek}{\omega_{\bk}} \right ) \ , \\
v_\bk^2 = \frac{1}{2} \left ( 1 - \frac{ \ek}{\omega_{\bk}} \right ) \ ,
\end{array}
\eeq 
we get 
\beq
A_{ij} = {\cal A}^0 ( \bR_{ij} ) + {\cal A}^1 (\bR_{ij} ) \ ,
\eeq with
\beq
\begin{array}{l}
{\cal A}^0 (\bR_{ij} ) \equiv i \om \int \frac{d^2 k}{(2 \pi)^2} \frac{e^{-i \bk \cdot
\bR_{ij} }} { ( i \om )^2 - \omega_{\bk}^2 } \ ,\\
{\cal A}^1 ( \bR_{ij}) \equiv  \int \frac{d^2 k}{(2 \pi)^2} \frac{\ek }{ (i \om)^2 -
\omega_{\bk}^2 } e^{-i \bk \cdot \bR_{ij} } \ . \end{array} 
\eeq
Similarly
\beq
C_{ij} = {\cal A}^0 ( \bR_{ij} ) - {\cal A}^1 (\bR_{ij} ) \ ,
\eeq and 
\bea
B_{ij} & = & \int \frac{d^2 k}{(2 \pi)^2} \frac{\dk}{( i \om )^2 - \omega_{\bk}^2 } e^{-i \bk
\cdot \bR_{ij} }  \nonumber \\
 & \equiv &  {\cal B}^1 ( \bR_{ji} ) \ .
\eea

\subsection{Evaluation of ${\cal A}^0 (\bR_{ji} ) $}
As we can see on figure~\ref{BZ} the spectrum has four nodes at the
points 
\beq
\begin{array}{cccc}
P_1 = (\frac{\pi}{2} , \frac{\pi}{2} )\ , \; \; \; \; & P_2 = ( -
\frac{\pi}{2}, \frac{\pi}{2} ) \ , \; \; \; \; & P_3 = ( - \frac{\pi}{2},
- \frac{\pi}{2} ) \ , \; \; \; \; & P_4 = ( \frac{\pi}{2}, -\frac{\pi}{2}
) \ .
\end{array} \eeq
Under the assumptions $\mu = 0 $ and $\Delta_0 = W$, we can linearize
the spectrum around each node in the following way:
\beq
\begin{array}{ll}
\bk^\prime \equiv ( \frac{\pi}{2}, \frac{\pi}{2} ) + \bk \ , \; \; \; \; &
\omega_{\bk^\prime}^2 = W^2 \left [ ( \cos k^\prime_x + \cos k^\prime_y
)^2 + ( \cos k^\prime_x - \cos k^\prime_y )^2 \right ] \ , \end{array} \eeq
and we get 
\beq \begin{array}{ll}
\omega_{\bk^\prime}^2 \simeq D^2 \ k^2  \ , \; \; \; \mbox{with} & \; \; D =
\sqrt{2} W  \ . \end{array} \eeq
Similarly,
\beq
\begin{array}{ll}
\ek^{\prime} \simeq W ( k_x + k_y ) \ , \\
\dk^{\prime} \simeq (k_x - k_y) \Delta_0 \ .
\end{array}
\eeq

In order to evaluate ${\cal A}^0 ( \bR )$ we divide the integral into
a sum of four integrals around each node :
\bea
{\cal A}^0 ( \bR ) & =  & \sum_{k^\prime \in BZ_1} i \om \frac{e^{-i
\bk^\prime \cdot \bR }} { ( i \om )^2 - \omega_{\bk^\prime}^2 } +  \sum_{k^\prime \in BZ_2} i \om \frac{e^{-i
\bk^\prime \cdot \bR }} { ( i \om )^2 - \omega_{\bk^\prime}^2 }\nonumber \\
&  + &  \sum_{k^\prime \in BZ_3} i \om \frac{e^{-i
\bk^\prime \cdot \bR }} { ( i \om )^2 - \omega_{\bk^\prime}^2 } +  \sum_{k^\prime \in BZ_4} i \om \frac{e^{-i
\bk^\prime \cdot \bR }} { ( i \om )^2 - \omega_{\bk^\prime}^2 } \ . \eea
In each term, the fact of developing around a particular node
gives a specific prefactor, so that we have
\beq
{\cal A}^0 ( \bR ) = {\cal F}^0 ( \bR ) \ (i \om ) \sum_k \frac{e^{-i
\bk \cdot \bR}}{(i \om)^2 - D^2 k^2 } \ ,
\eeq
with
\beq
\begin{array}{ll}
& {\cal F}^0 ( \bR ) = e^{-i \frac{\pi}{2} (R_x +
R_y)} + e^{-i \frac{\pi}{2} ( R_x - R_y )} + e^{ i \frac{\pi}{2} (R_x+ R_y ) } +
e^{i \frac{\pi}{ 2} (R_x - R_y)} \ . \end{array}
\eeq Thus
\beq
{\cal F}^0 (\bR ) = 2 \cos \frac{\pi}{ 2} (R_x + R_y ) + 2 \cos \frac{ \pi}{2} ( R_x
- R_y) \ . \eeq
Now calling $\theta$ the angle between $\bk$ and $\bR$ (cf. figure~\ref{angles}),
\bea
\label{omn}
{\cal A}^0 (\bR ) & = & {\cal F}^0 ( \bR ) (i \om ) \int_0^1 \frac{k \ dk
}{(2 \pi)^2} \int_0^{2 \pi} d \theta \frac{e^{-i k R \cos \theta}}{ (i
\om)^2 - D^2 k^2} \nonumber \\
 & = & - {\cal F}^0 ( \bR ) \frac{( i \om)}{ 2 \pi D^2} \int_0^1 k \ dk
 \frac{ J_0 ( k R) }{ (  \om  / D )^2 + k^2 } \nonumber \\
 & = & - {\cal F}^0 ( \bR ) \frac{i \om}{2 \pi D^2} K_0 \left ( R |
 \om | / D \right ) \ ,
\eea where $K_0$ is the Bessel function of rank zero. Note that we
 have defined $|\om| = \sqrt{- (i \om)^2}$.

\subsection{Calculation of ${\cal A}^1( \bR )$}
\begin{figure}[here]
\epsfxsize= 0.4 \textwidth
% ***********For one column  ********************
%\epsfxsize=7.0in 
% ***********************************8
\centerline{\epsfbox{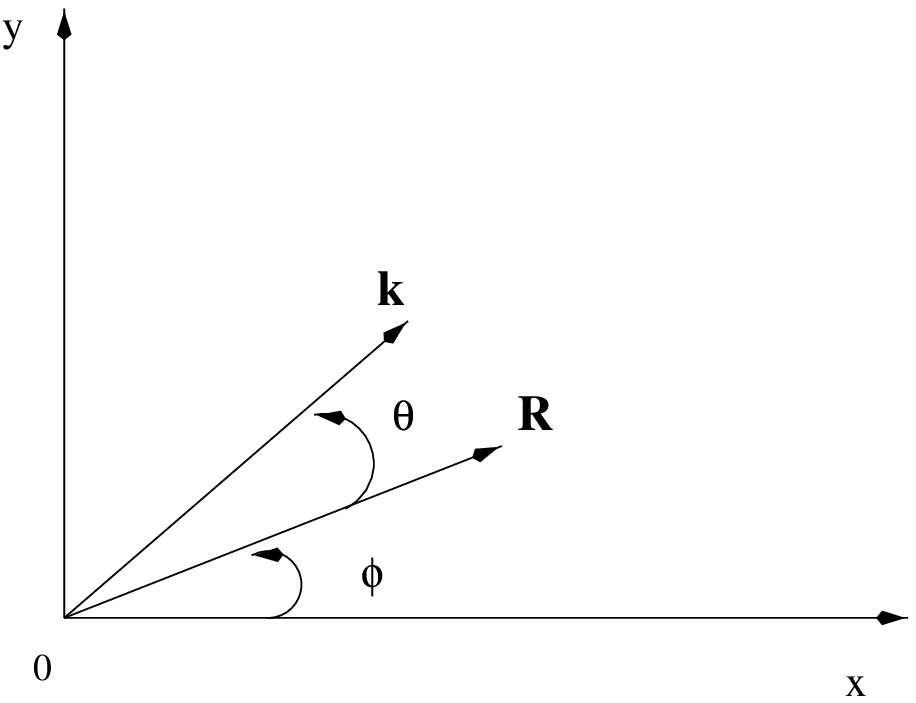}}
\vskip 0.1truein
\protect\caption{Angles between $\bR$ and $ \vec{x}$ and between $\bk$
and $ \bR$.}
\label{angles}
\end{figure}

As previously, we can decompose the summation in the Brillouin zone
into four parts:
\bea
\label{eqa1}
{\cal A}^1 ( \bR ) & =  & \sum_{k^\prime \in BZ_1} 
\frac{\varepsilon_{k^\prime} \ e^{-i
\bk^\prime \cdot \bR }} 
{ ( i \om )^2 - \omega_{\bk^\prime}^2 } +  \sum_{k^\prime
\in BZ_2}  \frac{\varepsilon_{k^\prime} \ e^{-i
\bk^\prime \cdot \bR }} { ( i \om )^2 - \omega_{\bk^\prime}^2 } \nonumber \\
&  + &  \sum_{k^\prime \in BZ_3}  \frac{ \varepsilon_{k^\prime} \ e^{-i
\bk^\prime \cdot \bR }} { ( i \om )^2 - \omega_{\bk^\prime}^2 } 
+  \sum_{k^\prime
\in BZ_4}  \frac{ \varepsilon_{k^\prime} \ e^{-i
\bk^\prime \cdot \bR }} { ( i \om )^2 - \omega_{\bk^\prime}^2 } \ . \eea
If we call $\theta$ the angle between $\bk$ and $\bR$ and $\varphi$
the angle between $\bR$ and the $x$-axis as represented on figure~\ref{angles}, the first term
in~(\ref{eqa1}) can be written
\bea
T_1 & \equiv & \sum_{k^\prime \in BZ_1} 
\frac{\varepsilon_{k^\prime} \ e^{-i
\bk^\prime \cdot \bR }} { ( i \om )^2 - \omega_{\bk^\prime}^2 } \nonumber \\
& = & e^{-i \frac{\pi}{ 2} (R_x + R_y)} \ \int_0^{2 \pi} d \theta \int_0^1 k \
dk \ \frac{D}{\sqrt{2} ( 2 \pi ) ^2} \frac{ ( k_x +k_y ) e^{-i k R \cos \theta}}{ (i
\om )^2 - D^2 k^2} \ .
\eea
Using
\beq
\begin{array}{l}
k_x = k ( \cos \theta  \cos \varphi - \sin \theta \sin \varphi ) \ ,\\
k_y = k ( \cos \theta \sin \varphi + \sin \theta \sin \varphi ) \ ,
\end{array}
\eeq 
we get
\bea
T_1 & = &  e^{-i \frac{\pi}{ 2} ( R_x + R_y )} ( \cos \varphi + \sin \varphi ) \
  \frac{D}{\sqrt{2} ( 2 \pi ^2)} \int_0^1 k^2 \ dk \int_0^{2 \pi} d
  \theta \frac{ ( \cos \theta + \sin \theta ) }{ (i \om )^2 - D^2 k^2 }
  e^{-i k R \cos \theta} \nonumber \\
& = & e^{-i \frac{\pi}{ 2} ( R_x + R_y )} ( \cos \varphi + \sin \varphi ) \
  \frac{ i D}{\sqrt{2}  2 \pi } \int_0^1 k^2 \ dk  \frac{ J_1( k R ) }{ (i \om )^2 - D^2 k^2 }
  e^{-i k R \cos \theta} \nonumber \\
& = & \frac{-i}{2 \sqrt{2} \pi} e^{-i \frac{ \pi}{ 2} ( R_x + R_y )} \frac{ \om }{D^2} K_1 \left ( \frac{ R | \om |}{D} \right) \ .
\eea Thus, after summing over the four nodes we get,
\beq
{\cal A}^1 ( \bR ) = \frac{{\cal F}^1 (\bR) }{2 \sqrt{2} \pi} \frac{
\om}{D^2} K_1   \left ( \frac{ R | \om |}{D} \right )  \ ,
\eeq with
\beq
{\cal F}^1 =2 \left [ \sin \frac{\pi}{2} ( R_x +R_y ) \left ( \cos \varphi +
\sin \varphi \right) + \sin \frac{\pi}{2} ( R_x - R_y ) \left ( \cos \varphi
- \sin \varphi \right) \right ] \ .
\eeq
The evaluation of ${\cal B}^1( \bR)$ is done in the same way as the
one of ${\cal A}^1( \bR)$.

	%\vspace{-.5  cm} % remove this before submission

\end{document}